\documentclass[aps,prd,10pt,twocolumn,preprintnumbers,superscriptaddress,bibnotes,nofootinbib,letterpaper,longbibliography]{revtex4-2}

\usepackage{hyperref}
\usepackage{graphicx}
\usepackage{amsfonts,amsmath,amssymb,bm}
\usepackage{physics}
\usepackage{svg}
\usepackage{array}
\usepackage[normalem]{ulem}
\usepackage[table]{xcolor} 

\hypersetup{
  colorlinks  = true,
  urlcolor    = blue,
  linkcolor   = blue,
  citecolor   = red
}

\newcommand{\orcid}[1]
{\begingroup
  \hypersetup{hidelinks}\href{https://orcid.org/#1}{\includegraphics[width=9pt]{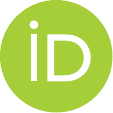}
} \endgroup}

\makeatletter
\newcounter{savesection}
\newcounter{apdxsection}
\renewcommand\appendix{\par
  \setcounter{savesection}{\value{section}}%
  \setcounter{section}{\value{apdxsection}}%
  \setcounter{subsection}{0}%
  \gdef\thesection{\@Alph\c@section}}
\newcommand\unappendix{\par
  \setcounter{apdxsection}{\value{section}}%
  \setcounter{section}{\value{savesection}}%
  \setcounter{subsection}{0}%
  \gdef\thesection{\@arabic\c@section}}
\makeatother

\usepackage[usenames,dvipsnames]{xcolor}

\begin{document}

\title{Towards First Detection of the Solar MSW Transition With JUNO}

\author{Obada Nairat \orcid{0000-0003-2019-9021}}
\email{nairat.2@osu.edu}
\affiliation{Center for Cosmology and AstroParticle Physics (CCAPP), \href{https://ror.org/00rs6vg23}{Ohio State University}, Columbus, OH 43210}
\affiliation{Department of Physics, \href{https://ror.org/00rs6vg23}{Ohio State University}, Columbus, OH 43210}

\author{John F. Beacom \orcid{0000-0002-0005-2631}}
\email{beacom.7@osu.edu}
\affiliation{Center for Cosmology and AstroParticle Physics (CCAPP), \href{https://ror.org/00rs6vg23}{Ohio State University}, Columbus, OH 43210}
\affiliation{Department of Physics, \href{https://ror.org/00rs6vg23}{Ohio State University}, Columbus, OH 43210}
\affiliation{Department of Astronomy, \href{https://ror.org/00rs6vg23}{Ohio State University}, Columbus, OH 43210} 

\author{Kevin J. Kelly \orcid{0000-0002-4892-2093}}
\email{kjkelly@tamu.edu}
\affiliation{Department of Physics and Astronomy, Mitchell Institute for Fundamental Physics and Astronomy, \href{https://ror.org/01red3556}{Texas A\&M University}, College Station, TX 77843}

\author{Shirley Weishi Li \orcid{0000-0002-2157-8982}}
\email{shirley.li@uci.edu}
\affiliation{Department of Physics and Astronomy, \href{https://ror.org/04gyf1771}{University of California}, Irvine, CA 92697}


\date{January 29, 2026}

\begin{abstract}
Matter-induced neutrino flavor mixing (the Mikheyev-Smirnov-Wolfenstein, or MSW, effect) is a central prediction of the neutrino mixing framework, but it has not been conclusively observed. Direct observation of the energy-dependent MSW transition in the solar electron-neutrino survival probability would solve this, but backgrounds have been prohibitive.  We show that our new technique for suppressing muon-induced spallation backgrounds will allow JUNO to measure the MSW transition at $>$4$\sigma$ significance in 10 years.  This would strongly support upcoming multi-\$1B next-generation experiments and their goals in cementing the neutrino mixing framework.
\end{abstract}

\preprint{UCI-HEP-TR-2025-26, MI-HET-872}

\maketitle


{\bf Introduction.---}
Neutrino flavor mixing in vacuum has been definitively established, showing that neutrinos have nonzero masses~\cite{Super-Kamiokande:1998kpq, Super-Kamiokande:2001ljr, Super-Kamiokande:2010tar, Super-Kamiokande:2016yck, SNO:2002tuh, SNO:2011hxd,  KamLAND:2002uet, KamLAND:2008dgz, Borexino:2011ufb, BOREXINO:2018ohr, DayaBay:2007fgu, DayaBay:2012fng}.  When neutrinos propagate through matter, their mixing can be substantially modified by the Mikheyev–Smirnov–Wolfenstein (MSW) effect~\cite{Wolfenstein:1977ue, Mikheyev:1985zog}, which arises from coherent forward scattering via the weak interactions.  Because these matter mixing effects depend on neutrino interactions at zero momentum transfer, they are highly sensitive to physics beyond the Standard Model (BSM)~\cite{Beacom:1999wx, Beacom:2002cb, Friedland:2004pp,  Berryman:2014qha, Maltoni:2015kca, Farzan:2017xzy, Huang:2018nxj, Proceedings:2019qno, Hostert:2020oui, Denton:2021vtf, Goldhagen:2021kxe,  Arguelles:2022tki, Coloma:2022umy, Coloma:2023ixt, Amaral:2023tbs, Denton:2024upc, Chattopadhyay:2025ccy, Gehrlein:2025isp}.  A quantitative experimental validation of matter mixing effects therefore remains an important goal.

However, no experiment has yet provided an unambiguous confirmation of matter mixing effects. For long-baseline and atmospheric experiments~\cite{T2K:2021xwb, NOvA:2021nfi, KM3Net:2016zxf, IceCubeCollaboration:2021euf, Super-Kamiokande:2023ahc, T2K:2025wet}, it is challenging in part because Earth’s column density is small. For solar experiments, the large and varying density of the Sun offers a powerful laboratory~\cite{Haxton:2012wfz, Gann:2021ndb, Xu:2022wcq}.  \textit{But while data on the electron-neutrino survival probability imply an MSW transition at a few MeV, it has never been directly observed}~\cite{Friedland:2004pp, Maltoni:2015kca, SNO:2011hxd, BOREXINO:2018ohr, Super-Kamiokande:2023jbt}.  Further, measurements of the solar day-night effect, which also probe the MSW theory, are similarly inconclusive~\cite{SNO:2011hxd, Super-Kamiokande:2023jbt}.

A precise understanding of matter mixing effects is crucial not only for fundamental physics, but also for practical reasons.  First, the Deep Underground Neutrino Observatory (DUNE) long-baseline experiment seeks to measure the neutrino mass ordering and the CP-violating phase~\cite{DUNE:2020lwj, DUNE:2020ypp}; these measurements either depend on matter effects or are vulnerable to degeneracies with BSM scenarios~\cite{Coloma:2015kiu, deGouvea:2015ndi, deGouvea:2016pom, Liao:2016orc}. (The Tokai to Hyper-Kamiokande, or T2HK, long-baseline experiment will also seek to measure the CP-violating phase, but under minimal matter effects~\cite{Hyper-Kamiokande:2018ofw}.)  Second, matter effects will be important for high-statistics measurements of atmospheric neutrinos in DUNE, Hyper-Kamiokande, and other experiments~\cite{ IceCubeCollaboration:2021euf, Coloma:2023ixt, Kelly:2021jfs, Hyper-Kamiokande:2022smq, KM3Net:2016zxf}. Third, matter effects are essential for interpreting observations of supernova neutrinos~\cite{Duan:2006an, Duan:2010bg, Chakraborty:2016lct, Capanema:2024hdm}.  

\begin{figure*}[t]
\includegraphics[width=0.90\textwidth]{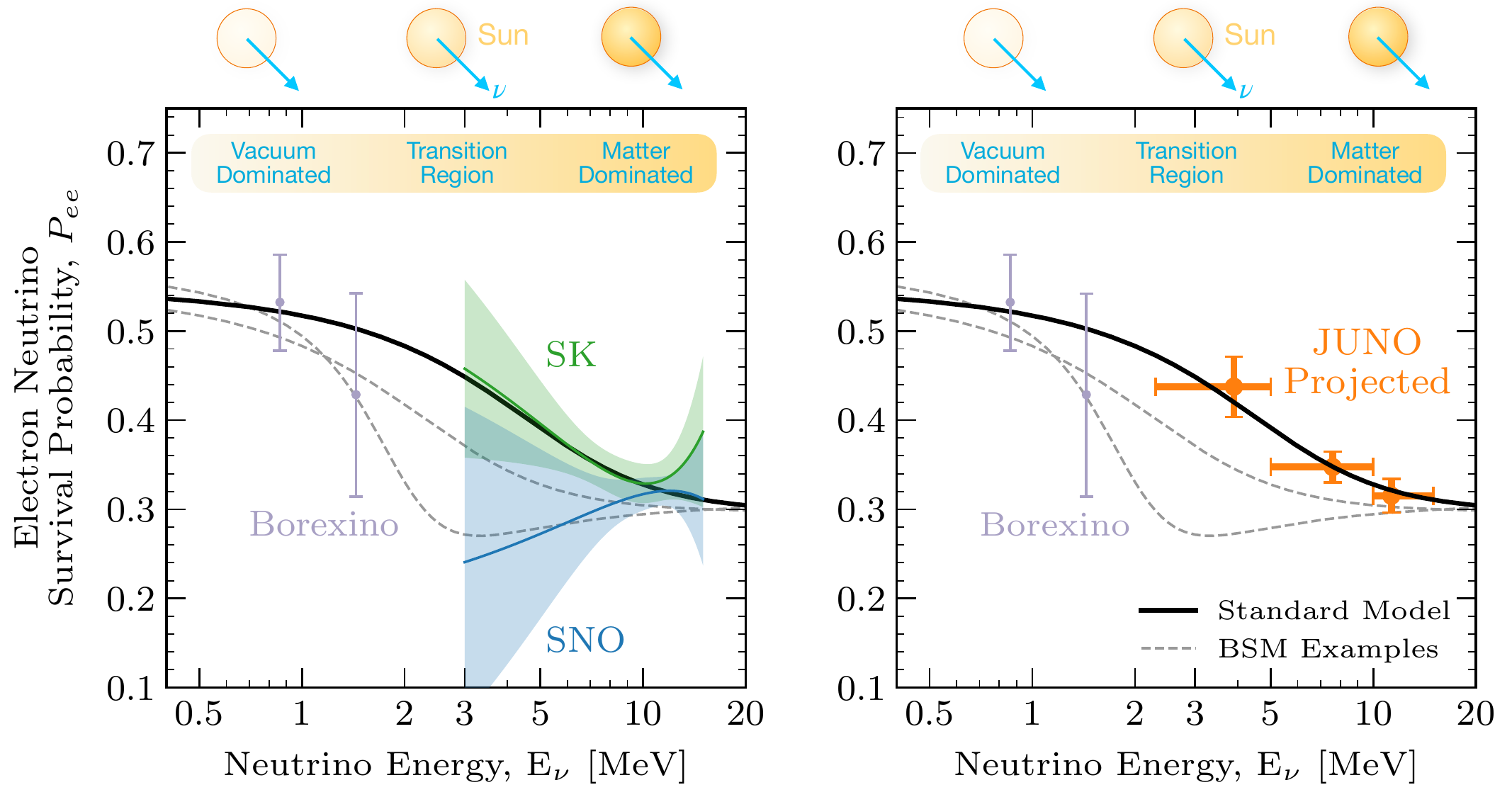}
\caption{Comparison of current~\cite{BOREXINO:2018ohr, Super-Kamiokande:2023jbt, SNO:2011hxd} and projected measurements (using our background-rejection techniques for JUNO) of the solar MSW transition. The black line shows the MSW prediction, while the gray dashed lines show example allowed NSI models. The bins shown for JUNO are a projection, assuming the Standard Model, of how they could go beyond the quadratic fits used by SK and SNO.  \textit{With our spallation cuts, JUNO can measure the MSW transition well.}}
\label{fig:SurvivalProb}
\vspace{-0.2cm}
\end{figure*}

In this \textit{Letter}, we show that the Jiangmen Underground Neutrino Observatory (JUNO)~\cite{JUNO:2015zny, JUNO:2020hqc, JUNO:2025gmd} --- designed primarily to measure reactor neutrinos --- can uniquely resolve this critical issue. In a companion paper~\cite{Nairat:2025sju}, we present a powerful new background rejection technique for JUNO measurements of \textit{solar neutrinos}. Exploiting this will give JUNO world-leading sensitivity to the MSW transition and will solidify the neutrino mixing framework. Figure~\ref{fig:SurvivalProb} contrasts the current experimental status versus our projected sensitivity for JUNO.

{\bf Review of the MSW effect.---} 
Neutrino mixing in vacuum can be described by a Hamiltonian evolution equation~\cite{Giunti:2007ry, Xu:2022wcq, Denton:2025jkt}.  This depends upon the mixing angles between the flavor and mass eigenstate bases ($\theta_{ij}$) and the mass-squared differences for the mass eigenstates ($\Delta m^2_{ji}$)~\cite{ParticleDataGroup:2024cfk}.  When neutrinos propagate in matter, the Hamiltonian is modified by adding the MSW term (``the matter potential") for electron neutrinos.  This term is $\propto G_F N_e$, where $G_F$ is the Fermi constant and $N_e$ is the electron density in the medium~\cite{Wolfenstein:1977ue, Mikheyev:1985zog,Giunti:2007ry, Maltoni:2015kca, Khan:2019doq, Coloma:2023ixt, Denton:2025jkt}.  For solar neutrinos of energy $E_\nu$, the electron-neutrino survival probability is~\cite{Parke:1986jy, Maltoni:2015kca}
\begin{align}
    P_{ee}(E_\nu) & \simeq \sin^4(\theta_{13}) + \cos^4(\theta_{13}) \, P_{ee}^{2\nu}(E_\nu), \\
    P_{ee}^{2\nu}(E_\nu) & = \frac{1}{2}[1 + \cos(2\theta_{12}) \langle \cos(2\theta_m)\rangle ],
    \label{eq:survival}
\end{align}
where $\cos(2\theta_m)$ refers to the matter mixing angle, and this term is averaged over the solar production region~\cite{Bahcall:2004pz, Bahcall:2005va, Vinyoles:2016djt, Acharya:2024lke, Denton:2025cbo, Zaidel:2025kdk}.  The energy dependence of Eq.~(\ref{eq:survival}) --- which provides a direct probe of matter mixing effects --- arises from $\theta_m$, which depends on the vacuum mixing angle and a dimensionless quantity $\propto G_F N_e E_\nu / \Delta m^2_{21}$. The angle $\theta_m$ is $\theta_{12}$ in vacuum (or low neutrino energies), $\pi/4$ at the MSW resonance density (or energy), and $\pi/2$ at high density (or high neutrino energies).

Figure~\ref{fig:SurvivalProb} shows the expected survival probability in the standard MSW framework using mixing parameters from the latest global fit~\cite{Esteban:2024eli} and recent JUNO measurements~\cite{JUNO:2025gmd}.  At the low energies probed by Borexino measurements (using $pp$, $pep$, and $^7$Be neutrinos)~\cite{BOREXINO:2018ohr}, the results are consistent with the vacuum limit of Eq.~(\ref{eq:survival}).  At the high energies probed by Super-Kamiokande (SK) and Sudbury Neutrino Observatory (SNO), the results are consistent with the high-density limit~\cite{SNO:2011hxd, Super-Kamiokande:2023jbt}.  In the few-MeV transition region, limited statistics lead to large uncertainties: SK data weakly favor an \textit{upturn} (at 1.2$\sigma$), while SNO data weakly favor a \textit{downturn} (at 0.7$\sigma$).  When taken together they somewhat favor an \textit{upturn} (at 2.1$\sigma$)~\cite{Super-Kamiokande:2016yck, Super-Kamiokande:2023jbt}.  These measurements are thus not yet sufficient to test matter mixing effects. The projected points for JUNO are discussed in a later section.

Figure~\ref{fig:SurvivalProb} also shows that the survival probability can be modified by BSM physics, where we focus on non-standard interactions (NSI) induced by new heavy mediators~\cite{Friedland:2004pp, Maltoni:2015kca, Farzan:2017xzy, Coloma:2017egw, Denton:2018xmq, Proceedings:2019qno, Arguelles:2022tki, Coloma:2022umy, Bernal:2022qba, Xu:2022wcq,  Denton:2022nol, Coloma:2023ixt, Denton:2024upc, Gehrlein:2025isp}.  Other possibilities include neutrino decay, sterile neutrinos, and large neutrino magnetic moments~\cite{Beacom:1999wx, Beacom:2002cb, Berryman:2014qha, Maltoni:2015kca, Huang:2018nxj, Goldhagen:2021kxe}. For NSI, the matter potential is modified by new interactions between neutrinos and fermions, adding a new term that is $\propto  G_F N_f \epsilon^f_{\alpha\beta} $, where $N_f$ is the density of the fermion $f$ in the medium and $\epsilon^f_{\alpha\beta}$ are dimensionless parameters that quantify the strength of the NSI between that fermion and the neutrino flavors $\nu_\alpha$ and $\nu_\beta$~\cite{Farzan:2017xzy, Coloma:2023ixt}.  We consider two example NSI models that are allowed at the $3\sigma$ level, corresponding to the parameters $(\epsilon_D,\epsilon_N)= (-0.20, -0.30)$ (bottom line) and $(-0.15,-0.12)$ (top line) as defined in Refs.~\cite{Coloma:2022umy, Amaral:2023tbs,  Shekar:2025xhx}. As an illustrative example, these parameter values can be generated through couplings to down quarks with $\epsilon^d_{ee} \sim 0.4$ and $\epsilon^d_{e\tau} \sim 0.1$. More generally, however, our analysis allows all relevant NSI couplings to electrons and quarks to vary.


{\bf Prospects for solar neutrinos in JUNO.---} 
Of present and planned detectors, JUNO has the best potential to make precise measurements of the MSW transition.  (For comparison, Hyper-Kamiokande expects to have a sensitivity of 3$\sigma$ in 10 years with an expected threshold of 4.5~MeV but may do better~\cite{Hyper-Kamiokande:2018ofw, Hyper-Kamiokande:2022smq}.) JUNO detects solar neutrinos through scattering with electrons, where the recoiling electron produces a prompt scintillation signal. The large scintillation light yield provides excellent energy resolution (3\%/$\sqrt{(E/{\rm MeV})}$) and enables a low detection threshold~\cite{JUNO:2020hqc, JUNO:2025fpc, JUNO:2025gmd}, both essential for detecting the MSW transition. With a 20-kton target mass, nearly 70 times larger than Borexino, JUNO will detect $^8$B neutrinos at a rate of $\sim$80~events per day through the elastic scattering channel~\cite{JUNO:2020hqc}. Although neutral-current and charged-current interactions on $^{13}$C also contribute~\cite{JUNO:2022jkf}, we focus here on elastic scattering because it provides the dominant contribution and has no kinematic threshold.

JUNO's analysis threshold for $^8$B solar neutrinos is set by detector backgrounds.  One important source is the natural radioactivity originating from contaminants in the scintillator or surrounding material. These backgrounds are effectively mitigated through multiple purification techniques and fiducial volume cuts~\cite{JUNO:2020hqc}. Initial performance results~\cite{JUNO:2025fpc} indicate that the concentrations of uranium and thorium in the detector are within the levels necessary to achieve a 2.3-MeV threshold, the lowest to date.  A somewhat high activity rate of polonium may be a challenge, but it should fall to the required level within roughly one year, based on its 138.4-day half-life. 

For the analysis window of 2.3--15~MeV, the dominant background is caused by the delayed decays of radioactive isotopes produced by cosmic-ray muons through spallation (nuclear breakup) processes~\cite{JUNO:2020hqc, Nairat:2025sju}.  The rate of these decays is $\sim$4000~events per day.  The standard approach to reducing spallation backgrounds is to veto all events within a cylindrical region following a cosmic-ray muon.  In Ref.~\cite{JUNO:2020hqc}, JUNO defined baseline cuts that would reduce spallation backgrounds to the sub-percent level, but at the cost of 50\% signal loss.


{\bf New technique to reduce backgrounds.---} 
For JUNO to achieve its full potential as a solar neutrino detector, improved background rejection techniques are required. In earlier papers on SK~\cite{Li:2014sea, Li:2015kpa, Li:2015lxa, Nairat:2024upg}, we showed that nearly all spallation isotopes are produced in muon-induced hadronic showers through the secondary particles associated with them, such as charged pions and neutrons. Importantly, hadronic showers are rare and are produced by $\lesssim$5\% of muons.  These considerations are general to other detectors~\cite{Zhu:2018rwc, Nairat:2025sju}. This motivates a strategy with aggressive cuts on hadronic showers and moderate cuts on electromagnetic showers.

\begin{figure}[t]
\includegraphics[width=0.99\columnwidth]{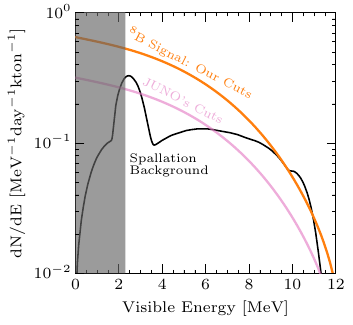}
\caption{The $^8$B signal spectrum at JUNO using our neutron-tagged cuts compared to using JUNO's cuts~\cite{JUNO:2020hqc}, both compared to the remaining spallation background spectrum.  \textit{With our spallation cuts, twice more signal would be retained.}}
\label{fig:ImprovedCuts}
\vspace{-0.2cm}
\end{figure}

In a companion paper~\cite{Nairat:2025sju}, we demonstrate that secondary neutrons are an efficient signature to tag hadronic showers in JUNO. These neutrons scatter down to thermal energies and capture on hydrogen, emitting a 2.2~MeV gamma ray, which JUNO detects with nearly perfect efficiency. For hadronic showers, we find an average yield of $\sim$14 neutrons, and at least one neutron 98\% of the time.  For hadronic showers, we apply cylindrical cuts with radii up to 6~m and times up to 40~s. For electromagnetic showers, we instead limit these to 3~m and 5~s. This strongly suppresses spallation backgrounds while keeping the deadtime (signal loss) minimal.

Figure~\ref{fig:ImprovedCuts} shows the increase in the $^8$B signal rate at JUNO using our neutron-tagged cuts (for the details of this figure, see Ref.~\cite{Nairat:2025sju}).  For the same background reduction efficiency, our method reduces the deadtime by a factor of about 4.5 compared to JUNO's baseline approach (from 62\% to 14\%), yielding a factor of 2.3 gain in exposure. This improvement in the signal-to-background ratio is critical for testing the MSW transition.


\begin{figure}[t]
\includegraphics[width=0.99\columnwidth]{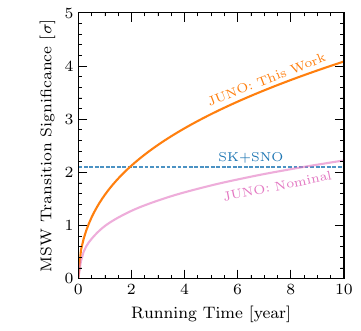}
\caption{Comparison of JUNO's median sensitivity to the solar MSW transition with our cuts and without, as well as to the combined SK+SNO result~\cite{Super-Kamiokande:2023jbt}.  \textit{With our spallation cuts, greater significance would be reached, and more quickly.}}
\label{fig:FlatSensitivity}
\vspace{-0.2cm}
\end{figure}

{\bf Probing the MSW transition.---} 
To test sensitivity to matter mixing effects, one must compare simulated data in JUNO under two hypotheses:
\begin{enumerate}
\vspace{-0.1cm}
    \item $P_{ee} =$~constant (but allowed to vary).
\vspace{-0.1cm}
    \item $P_{ee}$ given by the MSW theory in Eq.~(\ref{eq:survival}).
\vspace{-0.1cm}
\end{enumerate}
With JUNO's baseline cuts, they could reject the constant-probability hypothesis at $2.7\sigma$ significance in 10 years~\cite{JUNO:2020hqc}, somewhat better than the present combined SK+SNO result of $2.1\sigma$.

We calculate the recoil electron spectrum for each case by convolving the $^8$B neutrino spectrum~\cite{Winter:2004kf} with the elastic neutrino-electron cross section, including radiative corrections~\cite{Bahcall:1995mm}, and the appropriate survival probability. For the total flux of $^8$B solar neutrinos, we adopt the value measured using the neutral current channel at SNO~\cite{SNO:2009uok}, to avoid any implicit assumptions about the MSW transition. We incorporate detector effects, including the energy resolution ($\sim$3\% at 1~MeV), energy-scale uncertainty ($\sim$1\%)~\cite{JUNO:2020hqc, JUNO:2025fpc}, and detection efficiency~\cite{JUNO:2020hqc}. We apply the fiducial volume and selection cuts detailed in Ref.~\cite{JUNO:2020hqc}, including the $^{208}$Tl background cut which introduces a $\sim$20\% signal loss in the 3--5~MeV energy range and $\sim$2\% elsewhere~\cite{JUNO:2020hqc}. For spallation backgrounds, we apply the cuts detailed in the companion paper~\cite{Nairat:2025sju}. We then simulate all detector backgrounds remaining after the cuts, including spallation isotopes, intrinsic radioactivities, and reactor neutrino elastic scattering~\cite{JUNO:2020hqc}. We take into account statistical uncertainties as well as systematic uncertainties on the flux~\cite{SNO:2009uok} and spectrum shape~\cite{Winter:2004kf}. As in Ref.~\cite{JUNO:2020hqc}, the uncertainties associated with the residual spallation backgrounds are estimated based on previous measurements in scintillator detectors~\cite{KamLAND:2011fld}, but these can be validated and reduced by \textit{in-situ} measurements. Our analysis benefits modestly from the updated $^8$B spectrum of Ref.~\cite{Winter:2004kf}, which has reduced shape uncertainties compared to the one adopted in Ref.~\cite{JUNO:2020hqc}.

To quantify JUNO’s discrimination power, we follow the approach of their Ref.~\cite{JUNO:2020hqc}.  Assuming the MSW case as the truth and using bins of width 0.1~MeV, we evaluate the hypothesis of the flat-probability case using a $\chi^2$ test, taking into account all uncertainties.  Our analysis is slightly more conservative than JUNO's, as we start at 2.3~MeV instead of 2.0~MeV to take into account a new spallation background we identified (see Ref.~\cite{Nairat:2025sju}).

Figure~\ref{fig:FlatSensitivity} shows that JUNO, with improved spallation cuts and assuming the Standard Model, can reject the flat survival probability hypothesis with $>$4$\sigma$ significance in 10~years.  Moreover, JUNO will surpass the current SK+SNO result in about two years, less than half the time needed with JUNO's baseline cuts. Of course, JUNO results may also favor BSM scenarios.  This highlights the significance of our new background rejection techniques~\cite{Nairat:2025sju}.  We expect that our results can be improved further by developing likelihood-based cuts and employing machine learning techniques~\cite{Li:2018rzw, Zhang:2024ezk}. 

Note that the JUNO sensitivity is 2.2$\sigma$ instead of the 2.7$\sigma$ noted in Ref.~\cite{JUNO:2020hqc}.  This reflects their updated muon rate~\cite{JUNO:2025fpc}, which is higher.  We have verified that we can recover their original result with the old muon rate.

The sensitivity to the solar MSW transition depends on the value of the solar mass splitting. For lower values of $\Delta m_{21}^2$, the transition occurs at lower energies, degrading the sensitivity to the upturn given the detection threshold. Assuming a lower true value of $\Delta m_{21}^2 = 6.1 \times 10^{-5} \, \rm{eV}^2$, which is slightly favored by solar neutrino data~\cite{Super-Kamiokande:2023jbt}, the sensitivity to the MSW transition at JUNO decreases to $3.4\sigma$.

In Fig.~\ref{fig:SurvivalProb}, we show JUNO's projected measurements of the survival probability using three broad energy intervals, with the values obtained through a simultaneous fit to the full recoil electron spectrum.  These are for illustration and are not used in our main calculations. Because the mapping between neutrino energy and recoil electron energy is not one-to-one, correlations between the values are unavoidable. To minimize those, we choose sufficiently wide energy intervals, such that the remaining correlation coefficients are small (below $\sim$0.2) and are included in our uncertainty evaluation.  Compared to arbitrary functional fits~\cite{SNO:2011hxd, Super-Kamiokande:2023jbt}, our approach is more physical; compared to using narrow energy intervals, the visual uncertainties are easier to understand.  Further study is needed to optimize our approach.


{\bf Implications for future experiments.---} 
Measurements of long-baseline neutrino flavor mixing aim to determine, among other goals, the neutrino mass ordering and the value of the CP-violating phase. At DUNE, the ability to perform these measurements is linked directly to the presence of matter effects.  If BSM physics alters neutrino propagation in matter, it could pose significant challenges for these measurements. For example, NSI can substantially degrade the mass-ordering sensitivity~\cite{Deepthi:2016erc, Liao:2016orc, Coloma:2017egw, Denton:2018xmq, Esteban:2019lfo, Capozzi:2019iqn, Denton:2022nol} or even lead to errant interpretations of CP violation in long-baseline neutrino mixing~\cite{deGouvea:2016pom, Ge:2016dlx, Denton:2020uda, Chatterjee:2020kkm, NOvA:2024lti}.

\begin{figure}[t]
\includegraphics[width=0.99\columnwidth]{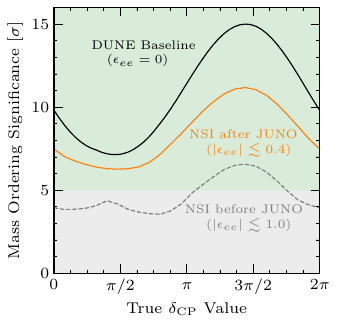}
\caption{DUNE sensitivity to the neutrino mass ordering with 6~years of data under different assumptions: Standard Model only, NSI within a currently allowed range, and NSI constrained by JUNO’s measurement of the MSW transition. \textit{JUNO's measurement of matter mixing effects will support DUNE's success.}}
\label{fig:DUNE_NSI}
\vspace{-0.2cm}
\end{figure}

To quantify the impact of matter-effect degeneracies on DUNE’s mass-ordering reach, we simulate expected event rates (in electron- and muon-neutrino samples) assuming 6~years of data collection --- equal running in neutrino and antineutrino modes --- with a detector fiducial mass of $20$~kton. Mixing parameters are fixed to the best-fit values from Ref.~\cite{Esteban:2024eli} (assuming the neutrino mass ordering is normal and with solar parameters updated from Ref.~\cite{JUNO:2025gmd}), with a varying assumed-true $\delta_{\rm CP}$. For the DUNE Baseline scenario, we allow $\theta_{23}$, $\delta_{\rm CP}$, and $\Delta m_{31}^2$ (fixing $\Delta m_{31}^2 < 0$) to vary, computing a $\chi^2$ test using bins of 250~MeV width in both $\nu_e$-appearance and $\nu_\mu$-disappearance channels. We then find a best-fit $\chi^2$ under the incorrect-ordering assumption, which is translated into the significance with which we can establish the mass ordering. We have also repeated this analysis assuming that the inverted mass ordering is true and find qualitatively similar results.

We consider representative NSI scenarios where the additional effective electron neutrino coupling is $\epsilon_{ee} = \sum_f \epsilon_{ee}^f N_f/N_e$. Current global constraints on $\epsilon_{ee}$ are model-dependent and vary depending on the mediator mass, but allow values roughly in the range $(-1,1)$ at the 3$\sigma$ level~\cite{Coloma:2022umy, Coloma:2023ixt}.  A precise measurement of the expected MSW transition in JUNO would restrict $\epsilon_{ee}$ to the range $(-0.2,0.2)$ at $1\sigma$ and $(-0.3,0.4)$ at $3\sigma$.

Figure~\ref{fig:DUNE_NSI} illustrates how JUNO’s measurement of the MSW transition can mitigate degeneracies that affect DUNE's sensitivity to the mass ordering. When all $\epsilon = 0$ (the Standard Model), DUNE will reach a significance of 7--15$\sigma$, depending on the value of the CP-violating phase, $\delta_{\rm CP}$.  Allowing NSI within currently allowed bounds introduces degeneracies that reduce this sensitivity to being below $5\sigma$ for the majority of $\delta_{\rm CP}$ values, which would undermine DUNE's goals.  
However, by independently probing matter mixing effects through solar neutrinos, JUNO will lift these degeneracies and restore DUNE’s mass-ordering sensitivity to more than $5\sigma$ for all $\delta_{\rm CP}$ within $\sim$5~years of exposure, compared to $\sim$10~years if the currently allowed NSI parameter space remains unconstrained. 

Beyond the mass-ordering sensitivity, NSI can also have a significant impact on measurements of CP violation. In particular, certain NSI scenarios can mimic CP-violating effects in long-baseline experiments, potentially leading to an apparent preference for nonzero $\delta_{\rm CP}$ even when the true value is CP conserving. For example, Ref.~\cite{deGouvea:2016pom} showed that an NSI coupling of $\epsilon_{ee}\sim0.7$ can generate a signal resembling maximal CP violation at DUNE when the true value is $\delta_{\rm CP}=0$. By independently constraining matter effects through precision measurements of the MSW transition, JUNO can exclude such large NSI couplings at the 3$\sigma$ level, providing a firm foundation for interpreting future measurements of leptonic CP violation.


{\bf Conclusions.---} 
We have shown that JUNO can provide a decisive experimental test of the MSW transition in solar neutrino mixing. By exploiting neutron tagging to muon-induced hadronic showers, spallation backgrounds in the critical 2.3--15 MeV energy range can be substantially reduced, enabling a measurement of the MSW transition with a significance $>$$4\sigma$ in 10 years.

Long-baseline accelerator experiments such as DUNE provide a concrete example of the importance of measuring the solar MSW transition, where JUNO alone can break degeneracies that limit their interpretation. In turn, this would provide a firm experimental foundation for next-generation measurements of the neutrino mass ordering and CP violation.  The JUNO measurement could be strengthened by combining with other neutrino mixing measurements in a global analysis.

The implications of the improved sensitivity enabled by our proposed spallation cuts extend beyond the measurement of the MSW transition. Another important open question in solar neutrino physics is the expected day--night asymmetry, caused by the regeneration of electron neutrinos as they propagate through the Earth. Because the sensitivity to this effect is currently limited primarily by statistics, the substantial exposure gain provided by our background-reduction strategy would also enhance JUNO’s ability to observe this effect.

More broadly, measuring the MSW transition would provide a model-independent calibration of matter effects, placing robust constraints on new physics that modifies neutrino propagation in matter. By doing so, a JUNO observation of the MSW transition would directly inform the interpretation of neutrino mixing across accelerator, atmospheric, and supernova experiments.


\medskip
{\bf Acknowledgments.---}
We thank the anonymous referees for their helpful and constructive feedback that greatly improved the manuscript. The work of ON and JFB was supported by National Science Foundation Grant No.\ PHY-2310018. KJK was supported in part by Department of Energy Award No.\ DESC0010813.
\clearpage
\appendix
\onecolumngrid
\section*{End Matter}
\twocolumngrid
\setcounter{equation}{0}
\renewcommand{\theequation}{E\arabic{equation}}
\setcounter{figure}{0}
\renewcommand{\thefigure}{E\arabic{figure}}

Here, we provide additional details that are helpful for reproducing our results, but which are not essential for the main arguments.  First, we summarize the NSI formalism.  Second, we show in more detail how NSI affects the survival probability of solar neutrinos, including a brief discussion of the LMA-Dark degeneracy. Finally, we describe in detail the analysis framework.

\smallskip

{\bf NSI formalism.---}
In the presence of non-standard neutrino interactions (NSI), neutrino propagation in matter is governed by an effective matter Hamiltonian that extends the Standard Model MSW potential. Here and in the main text, we closely follow Refs.~\cite{Coloma:2022umy, Coloma:2023ixt, Amaral:2023tbs, Shekar:2025xhx}, focusing only on vector NSI.  In the flavor basis, this Hamiltonian can be written as
\begin{equation}
    H_{\text{mat}}= \sqrt{2} G_F N_e
    \begin{pmatrix}
        1+\epsilon_{ee} & \epsilon_{e\mu} & \epsilon_{e\tau} \\
        \epsilon_{e\mu}^* & \epsilon_{\mu\mu} & \epsilon_{\mu\tau} \\
        \epsilon_{e\tau}^* & \epsilon_{\mu\tau}^* & \epsilon_{\tau\tau}
    \end{pmatrix},
\end{equation}
where the dimensionless parameters $\epsilon_{\alpha\beta}$ encode the strength of NSI relative to the Fermi constant.

In electrically neutral matter composed of electrons, protons, and neutrons, the effective NSI couplings $\epsilon_{\alpha\beta}$ can be expressed in terms of the fundamental NSI couplings to electrons, up quarks, and down quarks as
\begin{equation}
    \epsilon_{\alpha\beta}
    =
    \epsilon_{\alpha\beta}^e
    + (2\epsilon_{\alpha\beta}^u + \epsilon_{\alpha\beta}^d)
    + \frac{N_n}{N_e}(\epsilon_{\alpha\beta}^u + 2\epsilon_{\alpha\beta}^d),
\end{equation}
where $N_e$ and $N_n$ are the electron and neutron number densities, respectively.

For terrestrial experiments, the neutron density is approximately equal to the electron density, leading to an effective coupling commonly written as
\begin{equation}
    \epsilon_{\alpha\beta}^{\oplus}
    \simeq
    \epsilon_{\alpha\beta}^e
    + 3\epsilon_{\alpha\beta}^u
    + 3\epsilon_{\alpha\beta}^d.
\end{equation}
In contrast, in the Sun, the neutron fraction varies significantly with radius, from $Y_n \equiv N_n/N_e \simeq 0.6$ in the core to $\simeq 0.1$ near the surface. The effective solar NSI coupling can therefore be written as
\begin{equation}
    \epsilon_{\alpha\beta}^{\odot}
    \simeq
    \epsilon_{\alpha\beta}^e
    + (2+Y_n)\epsilon_{\alpha\beta}^u
    + (1+2Y_n)\epsilon_{\alpha\beta}^d.
\end{equation}
For our calculations in the main text, we use the neutron fraction profile from standard solar models~\cite{Bahcall:2004pz, Bahcall:2005va} to evaluate the radial dependence of $\epsilon_{\alpha\beta}^{\odot}$.

For solar neutrinos, the dynamics can be well described in the two-flavor limit. Then the effective $2\times2$ matter Hamiltonian takes the form~\cite{Coloma:2022umy, Amaral:2023tbs,  Shekar:2025xhx}
\begin{equation}
    H_{\text{mat}}\simeq \sqrt{2} G_F N_e
    \begin{pmatrix}
        c_{13}^2-\epsilon_D & \epsilon_N \\
        \epsilon_N^* & \epsilon_D
    \end{pmatrix},
\end{equation}
where the diagonal and off-diagonal NSI parameters, $\epsilon_D$ and $\epsilon_N$, are related to the effective couplings $\epsilon_{\alpha\beta}$ by
\begin{equation}
\begin{split}
    \epsilon_D =\;&
    c_{13}s_{13}\,\text{Re}(s_{23}\epsilon_{e\mu}+c_{23}\epsilon_{e\tau})\\
    &-(1+s_{13}^2)c_{23}s_{23}\,\text{Re}(\epsilon_{\mu\tau}) \\
    &-\frac{c_{13}^2}{2}(\epsilon_{ee}-\epsilon_{\mu\mu})
    +\frac{s_{23}^2-s_{13}^2c_{23}^2}{2}(\epsilon_{\tau\tau}-\epsilon_{\mu\mu}),
\end{split}
\label{eq:Eps_D}
\end{equation}
\begin{equation}
\begin{split}
    \epsilon_N =\;&
    c_{13}(c_{23}\epsilon_{e\mu}-s_{23}\epsilon_{e\tau}) \\
    &+ s_{13}\left[
        s_{23}^2\epsilon_{\mu\tau}
        -c_{23}^2\epsilon_{\mu\tau}^*
        +c_{23}s_{23}(\epsilon_{\tau\tau}-\epsilon_{\mu\mu})
    \right].
\end{split}
\label{eq:Eps_N}
\end{equation}
where $c_{ij} = \cos(\theta_{ij})$ and $s_{ij} = \sin(\theta_{ij})$.

\smallskip

{\bf NSI effects on the survival probability.---}
In the standard MSW framework, the effective matter mixing angle appearing in the Parke formula [Eq.~(\ref{eq:survival})] for the survival probability is
\begin{equation}
    \cos(2\theta_m)
    =
    \frac{
        \cos(2\theta_{12}) - \frac{A_{\rm CC}}{\Delta m_{21}^2} c_{13}^2
    }{
        \sqrt{
            \sin^2(2\theta_{12})
            +\left(
                \cos(2\theta_{12}) - \frac{A_{\rm CC}}{\Delta m_{21}^2} c_{13}^2
            \right)^2
        }
    },
\label{Eq:MatterMixingAngle}
\end{equation}
where $A_{\rm CC} = 2\sqrt{2}G_F N_e E_\nu$. At low neutrino energies or matter densities ($A_{\rm CC}/{\Delta m_{21}^2} \to 0$), the survival probability approaches a constant value, and at high neutrino energies or matter densities ($A_{\rm CC}/{\Delta m_{21}^2} \gg 1$), it approaches a different constant value.  In between, the energy dependence of $\theta_m$ --- as it varies from $\theta_{12}$ to $\pi/2$ --- produces the characteristic MSW transition.

In the presence of NSI, the terms in the numerator and denominator of Eq.~(\ref{Eq:MatterMixingAngle}) are modified as~\cite{Amaral:2023tbs,  Shekar:2025xhx}
\begin{align}
  \cos(2\theta_{12}) \to \cos(2\theta_{12})+\frac{2A_{\rm CC}}{\Delta m_{21}^2}\epsilon_D,\\
%
 \sin(2\theta_{12}) \to \sin(2\theta_{12})  +\frac{2A_{\rm CC}}{\Delta m_{21}^2}\epsilon_N
\label{eq:sin(2theta)}
\end{align}
%
\begin{figure*}[t]
    \centering
    \includegraphics[width=0.92\textwidth]{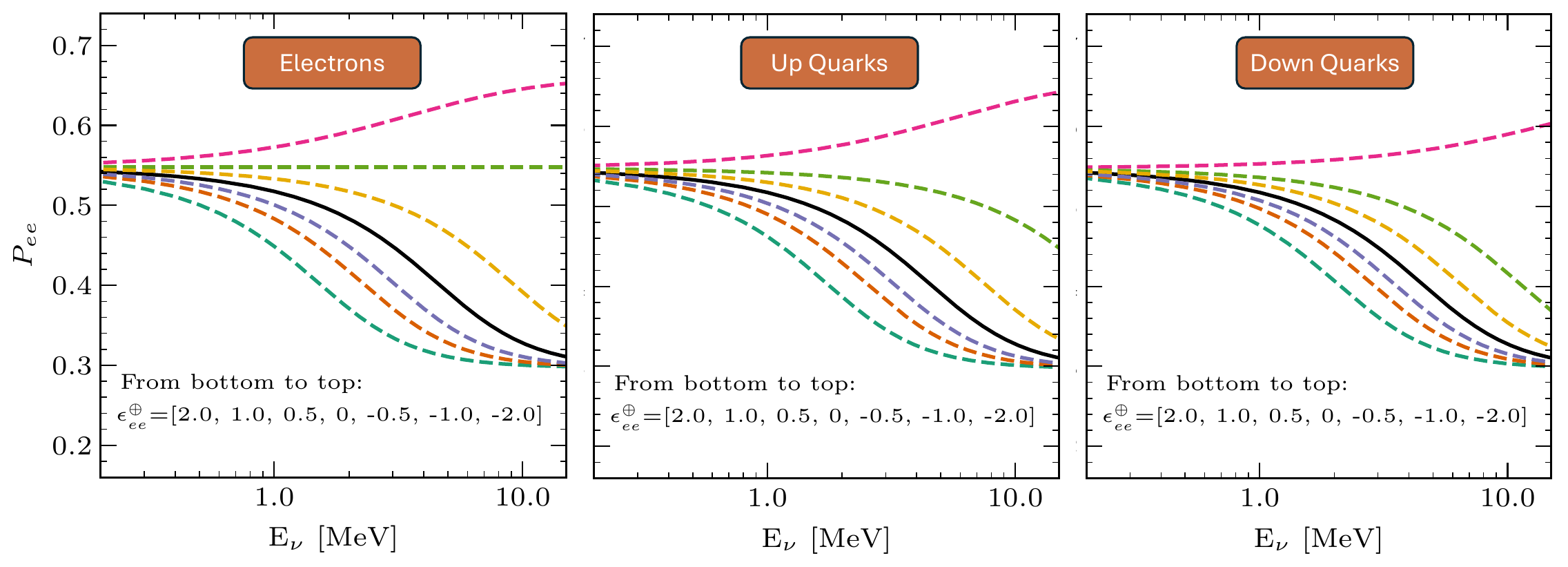}
    \caption{\vspace{-0.25cm}Solar neutrino survival probability in the presence of NSI couplings with different fermions.}
    \label{fig:A1}
    \vspace{-0.2cm}
\end{figure*}

Figure~\ref{fig:A1} illustrates the resulting distortions in the survival probability for representative values of $\epsilon_{ee}^{\oplus}$. The three panels correspond to scenarios where $\epsilon_{ee}^{\oplus}$ arises entirely from couplings to electrons, up quarks, or down quarks. The impact is largest for the case of electron couplings, for which the mapping between terrestrial and solar NSI parameters is unity, followed by the up and down-quark cases, where the proportionality factors are reduced by $(2+Y_n)/3$ and $(1+2Y_n)/3$, respectively.

The LMA-Dark degeneracy is a special but interesting NSI case~\cite{Coloma:2017egw, Denton:2018xmq, Denton:2022nol, Coloma:2023ixt, Chattopadhyay:2025ccy}.  Equations~(\ref{eq:Eps_D}--\ref{eq:sin(2theta)}) exhibit an exact degeneracy when $\epsilon_{ee}-\epsilon_{\mu\mu}=-2$, $\epsilon_{\mu\mu}=\epsilon_{\tau\tau}$, all off-diagonal NSI parameters vanish, and the sign of $\Delta m_{21}^2$ is reversed (equivalently, when $\theta_{12}$ lies in the second octant~\cite{Denton:2021vtf, Gehrlein:2025isp}). The LMA-Dark case thus cannot be separated from the standard model case using oscillation data alone.  Doing so requires complementary constraints from scattering measurements. In the electron sector, this parameter space is strongly constrained by neutrino–electron and coherent elastic neutrino–nucleus scattering data for a wide range of mediator masses~\cite{Coloma:2017egw, Denton:2018xmq, Denton:2022nol}.  In the near future, these constraints will be further strengthened by measurements of the solar neutrino fog at dark matter experiments~\cite{Gehrlein:2025isp}.

The importance of using scattering data extends beyond the LMA-Dark case alone.  In our calculations, we focus on the impact of NSI on neutrino propagation, which is independent of the mediator mass. For heavy mediators, NSI can also modify the neutrino–electron scattering cross section, providing an additional source of sensitivity. In current Borexino data, the NSI constraints are driven primarily by scattering effects because of the lack of data in the MSW transition region~\cite{Coloma:2022umy}. JUNO, by combining precision scattering measurements with its unique sensitivity to the MSW transition, will be able to place stronger and complementary constraints on NSI.
\smallskip

{\bf Analysis details.---}
In evaluating JUNO’s sensitivity to the solar MSW transition, we closely follow the statistical framework in Refs.~\cite{JUNO:2020hqc, Super-Kamiokande:2023jbt}, with some important updates motivated by recent simulations and detector performance studies.

To quantify the sensitivity, a $\chi^2$ function is defined as
\begin{equation}
    \chi^2 = \chi^2_{\rm data} + \sum_j \xi_j^2 /\sigma_j^2,
\end{equation}
where the index $j$ runs over the systematic uncertainties, $\xi_j$ are the corresponding nuisance parameters, and $\sigma_j$ are their errors, assumed to be Gaussian. For the data term, we use the Poisson definition,
\begin{equation}
    \chi_{\rm data}^2 = 2\sum_{i}N^i_{\rm{flat}} - N^i_{\rm MSW} +  N^i_{\rm MSW} \cdot\log\frac{ N^i_{\rm MSW}}{ N^i_{\rm flat}},
\end{equation}
where $N^i_{\rm{flat}}$ and $N^i_{\rm MSW}$ are the predicted event counts in the $i$th energy bin assuming, respectively, a flat (energy-independent) mixing probability and the standard MSW prediction. In both cases, signal and background contributions are included.

A key difference relative to Ref.~\cite{JUNO:2020hqc} is that we increase the analysis threshold from 2.0~MeV to 2.3~MeV. As discussed in detail in the companion paper~\cite{Nairat:2025sju}, updated \texttt{FLUKA} simulations indicate that the production yield of $^{13}$N is larger than previously estimated. This increase might be due to improved treatment of light-ion transport and the inclusion of additional production channels, such as ($^2$H,n) reactions. Because $^{13}$N has a long lifetime and a beta-decay endpoint at 2.2~MeV, it is particularly challenging to reduce using timing veto techniques. Raising the threshold to 2.3~MeV ensures complete removal of this background after accounting for energy smearing. While this choice leads to a modest reduction in sensitivity, it is essential to achieve a signal-dominated analysis.

The rates of residual spallation backgrounds are a major systematic uncertainty, due to the limited experimental measurements and significant hadronic uncertainties on the isotope yields, seen, e.g., in the different results from \texttt{Geant4} and \texttt{FLUKA}. In Ref.~\cite{JUNO:2020hqc}, isotope yields were estimated by scaling measurements from previous experiments, assuming a cross section proportional to the average muon energy.  (A more physical and accurate approach is to scale with shower energy~\cite{Li:2015kpa}.)  That approach leads to lower yields than those obtained in our \texttt{FLUKA} simulations. To remain conservative, for any isotope where the two approaches disagree, we adopt the larger predicted yield and further multiply it by a factor of two, which is the typical level of hadronic uncertainty.

For all remaining background components, we adopt the rates and uncertainties from Ref.~\cite{JUNO:2020hqc}, except in cases where updated measurements are available from JUNO’s recent results and initial detector performance studies~\cite{JUNO:2025fpc, JUNO:2025gmd}. In those instances, we use the experimentally measured values and their associated uncertainties. Altogether, this results in approximately $5\times10^4$ background events in a 10-year exposure, compared to about $3\times10^4$ events in Ref.~\cite{JUNO:2020hqc}, providing a conservative estimate of JUNO’s sensitivity. Ultimately, these yields and their uncertainties should be measured \textit{in-situ}.

Systematic uncertainties associated with the energy scale and the shape of the $^8$B neutrino spectrum are treated following the procedure of Ref.~\cite{JUNO:2020hqc}, by introducing energy-dependent shift parameters applied to the predicted spectra. The detector energy resolution is taken to be 3\% at 1~MeV. We have verified that increasing the resolution to 5\% has a negligible impact on the main results presented in this work.

\bibliography{upturn}


\end{document}